\documentclass[12pt,fleqn]{article}
\usepackage{amssymb}
\newcommand{\vek}[1]{\mathrm{\bf #1}}

\newcommand{\be}{\begin{equation}}
\newcommand{\ee}{\end{equation}}

\begin{document}
\hspace*{3cm}
\begin{center}
{\bf{\Large 
On the Newtonian Limit in Gravity
Models with Inverse Powers of $R$}}\\[1cm]
{\sc Rainer Dick}\\[5mm]
Department of Physics and Engineering Physics,\\
University of Saskatchewan, Saskatoon, Canada SK S7N 5E2\\[5mm]
\end{center}

\hspace*{0.49in}
\begin{minipage}{5in}{
{\bf Abstract:}
I reconsider the problem of the Newtonian limit
in nonlinear gravity models in the light of recently
proposed models 
$\mathcal{L}_{grav}\sim\sqrt{-g}f(R)$
with inverse powers of $R$.\\
Expansion around a maximally symmetric
local background with curvature scalar $R_0>0$ 
gives the correct Newtonian limit
on length scales $\ll R_0^{-1/2}$
if the gravitational Lagrangian
$\sqrt{-g}f(R)$
satisfies $|f(R_0)f''(R_0)|\ll 1$,
and I propose two models with $f''(R_0)=0$.
}
\end{minipage}

\section{Introduction}\label{sec:intro}

The need for an effective or genuine
cosmological constant to explain the 
faster than expected cosmological expansion in our epoch
has generated a lot of activity on scalar field ("quint\-essence")
models, where the potential energy
or an otherwise
anomalous dispersion relation of the quintessence
accelerates the expansion.

On the other hand, it is known that curvature 
terms can also accelerate the expansion of the universe
\cite{star,kerner,barrow,HJS0,HJS1,schmidtfp},
and the application of this mechanism to explain the 
current expansion rate has been denoted as curvature quintessence
\cite{capozziello}. While this mechanism usually relies on higher
order curvature terms, it has also been noticed recently
that inclusion of an $R^{-1}$ term in the gravitational Lagrangian
would yield a scale factor $a(t)\propto t^2$
\cite{capozziello,CCT,carroll}.

The model proposed recently by Carroll {\it et al.} (CDTT),
$\mathcal{L}\sim R-(\mu^4/R)$ \cite{carroll}, fits
into the framework of the so-called nonlinear gravity (NLG)
models
\begin{equation}\label{eq:action}
\mathcal{L}=\frac{M^2}{2}\sqrt{-g}f(R)+\mathcal{L}_{matter},
\end{equation}
see \cite{schmidtfp,magnano} and references there, and 
for brevity I denote models with $f(0)=\infty$ 
as {\it singular NLG models} in the sequel.

The generalized Einstein equations following from (\ref{eq:action})
are
\begin{eqnarray}
\tilde{G}_{\mu\nu}&\equiv&
f'(R)R_{\mu\nu}-\frac{1}{2}f(R)g_{\mu\nu}
-\nabla_\mu\nabla_\nu f'(R)
+g_{\mu\nu}\nabla^2 f'(R) \nonumber
\\
\label{eq:motion}
&=&\frac{1}{M^2}T_{\mu\nu},
\end{eqnarray}
and it is readily verified
that $\nabla^\mu\tilde{G}_{\mu\nu}\equiv 0$. NLG theories
usually assume\footnote{
I follow the MTW conventions \cite{MTW}
for the signature
of the metric and the definition of the Ricci tensor:
\[
R_{\mu\nu}=R^\sigma{}_{\mu\sigma\nu}
=\partial_\sigma\Gamma^\sigma{}_{\mu\nu}
-\partial_\nu\Gamma^\sigma{}_{\mu\sigma}
+\Gamma^\sigma{}_{\rho\sigma}\Gamma^\rho{}_{\mu\nu}
-\Gamma^\sigma{}_{\rho\nu}\Gamma^\rho{}_{\mu\sigma}.
\]
It is useful to keep that in mind when comparing with the literature
on regular NLG models, because the relative signs
between even and odd powers of $R$ depend on these conventions.
}
 $f(R)=R+6\ell^2 R^2+\mathcal{O}(R^3)$, whence
(\ref{eq:motion}) admits flat Minkowski space as a maximally symmetric
vacuum solution, and the Newtonian limit proceeds 
as in Einstein gravity, with additional
Yukawa terms in the gravitational potential \cite{sexl,stelle,HJS3,teyssandier}. 
Suppression of the Yukawa terms at macroscopic
distances leaves only the
the leading $1/r$ term, and one finds
that $M=M_{Pl}\equiv (8\pi G_N)^{-1/2}$ is the reduced Planck mass
as in Einstein gravity.

However, the model proposed in \cite{carroll} 
does not allow flat Minkowski space
as a solution, and 
the problem of the Newtonian limit is
more intricate\footnote{Capozziello {\it et al.} had noticed that
$f(R)=R^{-1}$ yields $a(t)\propto t^2$ \cite{capozziello,CCT}, but did not
further pursue this model. $f(R)=R^{-1}$ would not have
a maximally symmetric vacuum solution.}. 
Intuitively one would expect that on length
scales much smaller than an intrinsic curvature scale
one should be able to recover the Newtonian limit,
but intuition can be deceiving, and it is known in 
the framework of regular NLG models that these models 
may not admit a consistent weak field approximation. 
Therefore I propose the following approach to study this
problem for singular NLG theories: Since our four-dimensional
spacetime locally admits a ten-dimensional group
of symmetry transformations, the Newtonian limit, if it exists,
should be recoverable from expansion around a maximally
symmetric local background geometry, which contrary to the regular case
now will have to correspond to a curvature scalar $R_0\neq 0$.
This will be used in
Sec. \ref{sec:exp} to demonstrate that existence of a weak field
approximation around a symmetric local background with Ricci
scalar $R_0>0$ can be achieved by imposing the condition
$f''(R_0)=0$ on the singular NLG models.
In these models $M$ is then related
to the reduced Planck mass through 
\[
M=M_{Pl}\left/\sqrt{f'(R_0)}\right. .
\]

However, before entering the discussion of the Newtonian limit in singular
NLG models,
I would like to revisit and slightly extend the evidence for accelerated
expansion in these models in Sec. \ref{sec:cos}.

\section{The cosmological behavior at late times}\label{sec:cos}

The cosmological evolution equations from (\ref{eq:motion}) are
quite complicated, but we can make a general statement about the late
time expansion behavior of singular NLG models.

Since the generalized Einstein equation (\ref{eq:motion}) still
implies energy-momentum conservation $\nabla^\mu T_{\mu\nu}=0$, 
the time evolution of the
scale factor $a(t)$ in a Friedmann model is still governed
by the generalized Friedmann equation 
$\delta\mathcal{L}/\delta g^{00}|_{FRW\,\, metric}=0$.

 For the spatially flat FRW metric
\[
ds^2=-dt^2+a^2(t)d\vek{x}^2
\]
the generalized Friedmann equation following
from (\ref{eq:motion}) is
\begin{eqnarray}
&&-3f'\left(6\frac{\ddot{a}}{a}+6\frac{\dot{a}^2}{a^2}\right)
\frac{\ddot{a}}{a}
+\frac{1}{2}f\left(6\frac{\ddot{a}}{a}+6\frac{\dot{a}^2}{a^2}\right)
+3\frac{\dot{a}}{a}\partial_0
f'\left(6\frac{\ddot{a}}{a}+6\frac{\dot{a}^2}{a^2}\right)
\nonumber
\\ \label{eq:fried}
&&=\frac{1}{M^2}\varrho,
\end{eqnarray}
with $T_{00}=\varrho$.

In general this will be a third order equation for the scale factor.
To analyze the late time behavior, we first assume that there is only
ordinary dust and radiation in $\varrho$, whence the energy density
can be neglected at late times for expanding solutions.

We then make a power law {\it ansatz} $a(t)\propto t^\alpha$,
which yields
\begin{eqnarray}
&&-3f'\left(\frac{6}{t^2}\alpha(2\alpha-1)\right)
\frac{\alpha(\alpha-1)}{t^2}
+\frac{1}{2}f\left(\frac{6}{t^2}\alpha(2\alpha-1)\right)
\nonumber
\\
\label{eq:fried2}
&&-36\frac{\alpha^2(2\alpha-1)}{t^4}
f''\left(\frac{6}{t^2}\alpha(2\alpha-1)\right)
=0.
\end{eqnarray}

If $R^{-n}$, $n>0$, is the leading order singularity in the
singular NLG model $f(R)$, then at late times the contribution
from this term will dominate all 3 terms in Eq. (\ref{eq:fried2}),
with the same time dependence $\propto t^{2n}$. This yields
the algebraic equation
\[
n\frac{\alpha-1}{2(2\alpha-1)}+\frac{1}{2}
-\frac{n(n+1)}{2\alpha-1}=0,
\]
which determines the expansion coefficient $\alpha$:
\begin{equation}\label{eq:alpha}
\alpha=\frac{2n^2+3n+1}{n+2}.
\end{equation}

This was found for $f(R)=R-\mu^{2n+2}R^{-n}$ in \cite{carroll}
through conformal transformation to a corresponding scalar quintessence
model, and the corresponding result for $f(R)=R^n$, $n>0$, is also
spelled out in \cite{CCCT}.

Note, however, that Eq. (\ref{eq:fried}) is also compatible with 
exponential expansion $a(t)\propto\exp(Ht)$ at late times
if $f(12H^2)=6H^2f'(12H^2)$ has a solution. Carroll {\it et al.} 
found in the metric formulation
of their model that power law expansion is dynamically preferred
\cite{carroll}. Vollick looked at the Palatini formalism in the
CDTT model and concluded that exponential expansion would arise
in that formulation \cite{dan}. Our use of a symmetric
local background geometry in the next section does not predicate the global
late time expansion, but only assumes that spacetime
should have maximal symmetry locally.

\section{Expansions around maximally symmetric vacua}\label{sec:exp}

In the spirit of the philosophy outlined in Sec. \ref{sec:intro}
we now assume that the Newtonian limit should be recoverable
through weak field expansion around a symmetric local background geometry:
The maximally symmetric vacuum solutions satisfy
\[
R_{\alpha\mu\beta\nu}
=\frac{R_0}{12}(
g_{\alpha\beta}g_{\mu\nu}
-g_{\alpha\nu} g_{\mu\beta}),
\]
\[
R_{\mu\nu}=\frac{R_0}{4}g_{\mu\nu},
\]
where the constant curvature scalar $R_0$ satisfies
\begin{equation}\label{eq:condr0}
f'(R_0)R_0=2f(R_0).
\end{equation}
In ordinary NLG theories this always permits $R_0=0$, but
 for singular NLG models this yields values $R_0\neq 0$,
and the challenge is to derive the Newtonian limit 
from the weak field expansion around the vacuum solution.

The first order expansion of Eq. (\ref{eq:motion}) around a
symmetric vacuum solution yields
\begin{eqnarray}
&&f'(R_0)\delta R_{\mu\nu}+\frac{1}{4}[f''(R_0)R_0-2f'(R_0)]
g_{\mu\nu}\delta R
\nonumber
\\ \label{eq:motionT}
&&-\frac{1}{2}f(R_0)\delta g_{\mu\nu}
-f''(R_0)(\nabla_\mu\nabla_\nu\delta R
-g_{\mu\nu}\nabla^2\delta R)
=\frac{1}{M^2}T_{\mu\nu}
\end{eqnarray}
or
\begin{eqnarray}
&&f'(R_0)\delta R_{\mu\nu}-\frac{1}{4}f''(R_0)R_0
g_{\mu\nu}\delta R
-\frac{1}{2}f(R_0)\left(\delta g_{\mu\nu}
-\frac{1}{2}g_{\mu\nu}\delta g\right)
\nonumber
\\ \label{eq:motion2}
&&-f''(R_0)\left(\nabla_\mu\nabla_\nu\delta R
+\frac{1}{2}g_{\mu\nu}\nabla^2\delta R\right)
=\frac{1}{M^2}\left(T_{\mu\nu}-\frac{1}{2}g_{\mu\nu}T
\right).
\end{eqnarray}

The first order variation of the Ricci tensor
is
\begin{eqnarray}
\delta R_{\mu\nu}&=&
\frac{1}{2}(\nabla_\mu\nabla^\sigma\delta g_{\sigma\nu}
+\nabla_\nu\nabla^\sigma\delta g_{\sigma\mu}
+\delta g_{\alpha\nu}R^\alpha{}_\mu
+\delta g_{\alpha\mu}R^\alpha{}_\nu)
\nonumber
\\
&&-\delta g_{\alpha\beta}R^\alpha{}_\mu{}^\beta{}_\nu
-\frac{1}{2}\nabla^2\delta g_{\mu\nu}
-\frac{1}{2}\nabla_\mu\nabla_\nu\delta g
\\ \label{eq:dR}
&=&
\frac{1}{2}(\nabla_\mu\nabla^\sigma\delta g_{\sigma\nu}
+\nabla_\nu\nabla^\sigma\delta g_{\sigma\mu})
+\frac{1}{3}R_0\delta g_{\mu\nu}
-\frac{1}{12}R_0 g_{\mu\nu}\delta g
\\ \nonumber
&&-\frac{1}{2}\nabla^2\delta g_{\mu\nu}
-\frac{1}{2}\nabla_\mu\nabla_\nu\delta g,
\end{eqnarray}

The mass term $f(R_0)\sim\mathcal{O}(R_0)$
vanishes in regular NLG theories, and
should also be negligible in singular
NLG theories at least up to length scales where
Newton's law has been verified, which implies that
$R_0\sim\mu^2$ must correspond to a small mass scale
$\mu$. We also note that the mass and derivative
terms following from Eqs. 
(\ref{eq:motion2}) and (\ref{eq:dR}) have the correct signs
for non-oscillatory attractive solutions if 
$R_0\ge 0$, $f'(R_0)>0$ ($\Rightarrow f(R_0)\ge 0$), 
and $f''(R_0)\ge 0$.

In regular NLG theories $f(R_0)=f(0)=0$,
$f'(R_0)=1$, and
$f''(R_0)=12\ell^2$ is
assumed to be very small, such that
the corresponding Yukawa terms are 
suppressed relative to the leading $1/r$
term at macroscopic distances.
On the other hand, every set of observational tests of Newton's
law can only cover a finite range of length scales.
Therefore one might be tempted to conclude that
very large $\ell$ is another possibility,
such that e.g. the Yukawa term $\exp(-r/\ell)/r$
from $f(R)=R+6\ell^2 R^2$
at observational distances also approximates a $1/r$ term
and only rescales the ratio between
$M$ and $M_{Pl}$ by a constant factor.

That this latter possibility is excluded in regular NLG theories
was noticed already by Pechlaner and Sexl: $f''(0)=12\ell^2$
has to be small for consistency of the weak field approximation,
because otherwise domination of the fourth order terms
would yield strong curvature on all length scales \cite{sexl}.

This reasoning carries over to the singular case, with minor 
modification:
Due to the presence of a small mass term the 
Newtonian potential, if it exists in the theory, will always 
come from a limit of Yukawa terms.
Yet we still have to confine the impact
from the fourth order terms to small $r$.
This will constrain the parameter space,
because 
in singular NLG theories $f'(R_0)\sim
\mathcal{O}(1)$, and $f''(R_0)\sim\mu^{-2}$
generically would
imply that the fourth order derivative terms
dominate the equation for $\delta g_{\mu\nu}$, thus spoiling the consistency
of the weak field approximation.

The need for suppression of
the fourth order terms can also be seen from the following simple example:
\[
\Delta U(\vek{r})-\mu^2 U(\vek{r})
-\frac{1}{2m^2}\Delta^2 U(\vek{r})=\frac{1}{2M^2}\delta(\vek{r})
\]
yields
\[
U(\vek{r})=-\frac{m}{8\pi M^2 r\sqrt{m^2-2\mu^2}}
\left[\exp(-k_- r)-\exp(-k_+ r)\right],
\]
with
\[
k^2_\pm=m^2\pm m\sqrt{m^2-2\mu^2}.
\]
This will give a Newtonian $1/r$ potential 
at distances $r\ll\mu^{-1}$
only if $\mu\ll m$:
\begin{eqnarray*}
U(\vek{r})&\approx& -\frac{1}{8\pi M^2 r}
\left[\exp(-\mu r)-\exp(-\sqrt{2}m r)\right]
\\
&\approx& -\frac{1}{8\pi M^2 r},
\quad m^{-1}\ll r\ll\mu^{-1}.
\end{eqnarray*}

In the terminology of the singular NLG models
this means that we need 
\[
|f(R_0)f''(R_0)|\ll 1, 
\]
while e.g. $f(R)=R-\mu^{2n+2}R^{-n}$ would yield
$|f(R_0)f''(R_0)|=n(n+1)^2/(n+2)^2$.

Therefore we either have to invoke a second small parameter in $f(R)$
such that {\it both} $f(R_0)$ and $f''(R_0)/r^4$ are small
relative to $f'(R_0)/r^2$ at the length scales of interest. 
Or, since $f(R_0)\neq 0$ 
by Eq. (\ref{eq:condr0}), we arrange $f(R)$ such that
the coefficient of $\mu^{-2}$ vanishes altogether,
i.e. by choosing our model such that
the solution of Eq. (\ref{eq:condr0}) satisfies
\begin{equation}\label{eq:condf2}
f''(R_0)=0.
\end{equation}
In that case the fourth order terms vanish 
in the weak field expansion and all the curvature contributions
to Eq. (\ref{eq:motion2}) are subleading, such that for $r\ll\mu^{-1}$
a flat {\it ansatz} can be used to determine the local potential at these
scales. In leading order this is then nothing but the ordinary Newtonian
limit at these scales, with the only modification that
\begin{equation}\label{eq:M}
M=(8\pi G_N f'(R_0))^{-1/2}
=M_{Pl}\left/\sqrt{f'(R_0)}\right. .
\end{equation}

\section{Two simple examples of singular NLG models}\label{sec:dmod}

{\bf 4.1} The criterion (\ref{eq:condf2})
is satisfied e.g. by
\begin{equation}\label{eq:newmod1}
\mathcal{L}=\frac{M^2}{2}\sqrt{-g}\left(
R+\frac{R^2}{9\mu^2}-\frac{3\mu^4}{R}
\right)
+\mathcal{L}_{matter}.
\end{equation}
This corresponds to
\[
R_0=3\mu^2
\]
and
\[
M=M_{Pl}\left/\sqrt{2}\right. ,
\]
and the power law for late time expansion 
would be the same as in the original CDTT model,
$a(t)\propto t^2$.\\[2mm]
{\bf 4.2} Another model that satisfies the criterion (\ref{eq:condf2})
is
\begin{equation}\label{eq:newmod2}
\mathcal{L}=\frac{M^2}{2}\sqrt{-g}\left(
R-15\frac{\mu^4}{R}+25\frac{\mu^6}{R^2}
\right)
+\mathcal{L}_{matter}.
\end{equation}
This yields
\[
R_0=5\mu^2
\]
and
\[
M=M_{Pl}\sqrt{\frac{5}{6}}.
\]
The $R^{-2}$ term accelerates the power law
expansion at late times to
$a(t)\propto t^{3.75}$.

\section{Conclusions}\label{sec:conc}

The problem of existence of a weak field expansion and the Newtonian
limit is more intricate in singular NLG models
than in regular NLG models, but can apparently
be solved.

Models can in particular be chosen to satisfy the constraint 
(\ref{eq:condf2}) to ensure consistency
of the weak field expansion at length scales $\ll R_0^{-1/2}$.

Two minimal extensions of the CDTT model
which satisfy this constraint
are given in Eqs. (\ref{eq:newmod1}) and (\ref{eq:newmod2}).\\[2mm]
{\bf Acknowledgement:} My research is supported through
the Natural Sciences and Engineering Research Council
of Canada.

\end{document}